\documentclass[aps,prb,preprint,superscriptaddress,showpacs]{revtex4}
\usepackage{graphicx}
\usepackage{bm}

\begin{document}
\title{Resonant inelastic x-ray scattering study of hole-doped manganites 
La$_{1-x}$Sr$_x$MnO$_3$ ($x=0.2$ and $0.4$)}
\author{K. Ishii}
\email{kenji@spring8.or.jp}
\affiliation{Synchrotron Radiation Research Center, Japan Atomic
Energy Research Institute, Hyogo 679-5148, Japan}
\author{T. Inami}
\affiliation{Synchrotron Radiation Research Center, Japan Atomic
Energy Research Institute, Hyogo 679-5148, Japan}
\author{K. Ohwada}
\affiliation{Synchrotron Radiation Research Center, Japan Atomic
Energy Research Institute, Hyogo 679-5148, Japan}
\author{K. Kuzushita}
\affiliation{Synchrotron Radiation Research Center, Japan Atomic
Energy Research Institute, Hyogo 679-5148, Japan}
\author{J. Mizuki}
\affiliation{Synchrotron Radiation Research Center, Japan Atomic
Energy Research Institute, Hyogo 679-5148, Japan}
\author{Y. Murakami}
\affiliation{Synchrotron Radiation Research Center, Japan Atomic
Energy Research Institute, Hyogo 679-5148, Japan}
\affiliation{Department of Physics, Tohoku University, Sendai
980-8578, Japan}
\author{S. Ishihara}
\affiliation{Department of Physics, Tohoku University, Sendai
980-8578, Japan}
\author{Y. Endoh}
\affiliation{Institute for Materials Research, Tohoku University,
Sendai 980-8577, Japan}
\affiliation{International Institute for Advanced Studies, Kizugawadai,
Kizu, Kyoto 619-0225, Japan}
\author{S. Maekawa}
\affiliation{Institute for Materials Research, Tohoku University,
Sendai 980-8577, Japan}
\author{K. Hirota}
\affiliation{The Institute for Solid State Physics, The University of
Tokyo, Chiba 277-8581, Japan}
\author{Y. Moritomo}
\affiliation{Department of Applied Physics, Nagoya University,
Nagoya 464-8603, Japan}
\date{\today}

\begin{abstract}
Electronic excitations near the Fermi energy in the hole doped
manganese oxides (La$_{1-x}$Sr$_x$MnO$_3$, $x=0.2$ and $0.4$) have
been elucidated by using the resonant inelastic x-ray scattering (RIXS)
method. A doping effect in the strongly
correlated electron systems has been observed for the first time. The scattering spectra
show that a salient peak appears in low energies indicating the
persistence of the Mott gap. At the same time, the energy gap is
partly filled by doping holes and the energy of the spectral weight
shifts toward lower energies. The excitation spectra show little
change in the momentum space as is in undoped LaMnO$_3$, but the
scattering intensities in the low energy excitations of $x=0.2$ are
anisotropic as well as temperature dependent, which indicates a
reminiscence of the orbital nature.
\end{abstract}


\maketitle

\section{Introduction}
Strongly correlated electron systems (SCES), in particular the
transition metal oxides, have attracted much attention for more than a
decade. They provide not only many novel physical properties, such as
high temperature superconductivity, enhanced anomalies in the
conductivity, etc.\ but the playground for the elucidation of the most
fundamental subject of many-body interactions of electrons as well as
the interplay of different freedoms of charge, spin, orbital and
lattice. Manganese perovskites often called as manganites now become
extremely important materials for this reason, in particular of the
colossal magneto-resistance (CMR) effect. The CMR effect has been
primarily discussed as the result of mutual interactions of various
freedoms and then important issues are much relied on the search of
the electronic structure in the manganites. More precisely, the effect
of the strong electron correlations and the interactions with spin and
orbitals determine the momentum dependence of the electron energies
near the Fermi energy which is modified from the typical Fermi liquid
picture in the textbook. Such electronic structure can be thoroughly
explored by the detailed experiments of searching the charge dynamics
in various manganites. In this respect, the resonant inelastic x-ray
scattering (RIXS) is an ideal tool for the elucidation of the electron
excitations, since it gives direct notion of momentum dependent
spectrum. There exist several reports of such studies since the
pioneering work on NiO by Kao et al.\cite{Kao1}. Recently the RIXS
measurements have been directed to both cuprates \cite{Hill1,
Abbamonte1, Hamalainen1, Hasan1, Hasan2, Kim1} and manganites
\cite{Inami2}. For the purpose of the present research, the RIXS study
in LaMnO$_3$ of the progenitor of the CMR compounds is briefly
summarized here\cite{Inami2}. Three salient peaks appear at the
transferred energies of 2.5, 8 and 11 eV by injecting photons of the
resonant energy corresponding to the Mn $K$ absorption edge. The
lowest excitation peak appeared at 2.5 eV was assigned to be the
excitations from the occupied effective lower Hubbard band (LHB) which
consists of hybridized O $2p$ and Mn $3d$ orbitals to the empty upper
Hubbard band (UHB) based on the theoretical calculation of the RIXS
process \cite{Kondo1}. The excited electron changes the character of
orbital from $3d_{3x^2-r^2}/3d_{3y^2-r^2}$ to
$3d_{y^2-z^2}/3d_{z^2-x^2}$.  Experimental facts of a weak momentum
dependence and the apparent polarization dependence of the scattering
intensity are consistent with the theoretical calculations. Two other
resonant peaks at higher transition energies were also assigned to be
charge transferred excitations from the inner O $2p$ band to the
partially filled Mn $3d$ band and the empty band of hybridized
$4s$/$4p$ orbital, respectively.

\begin{figure}
\includegraphics[scale=0.45]{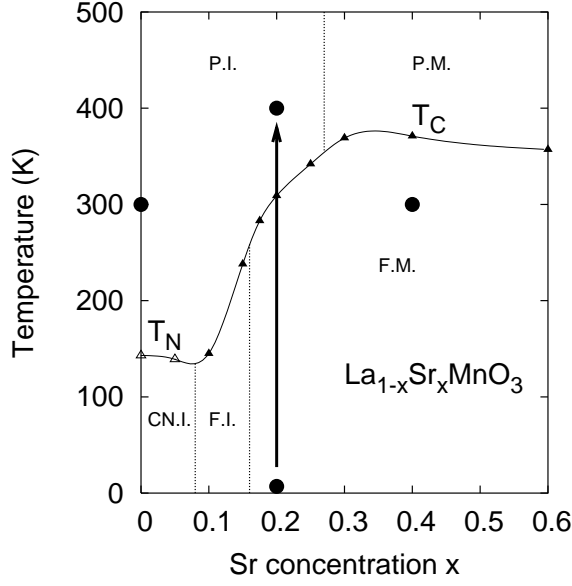}
\caption{Electronic phase diagram of La$_{1-x}$Sr$_x$MnO$_3$ for the
Sr concentration ($x$) and temperature ($T$) from a
reference\cite{Urushibara1}. The filled circles and the arrow indicate
the observed points in Fig.\ \ref{fig:xdep} and \ref{fig:x02tdep},
respectively.}
\label{fig:phase}
\end{figure}

Our primary goal of the present RIXS studies is to observe the
excitation spectra from the metallic phase of the transition metal
oxides, because novel properties appear in the vicinity of the metal
insulator transition (MIT) by the carrier doping into the Mott
insulators. No publications on the RIXS studies in the metallic phases
exist so far, though we hear numerous experimental efforts to
elucidate the nature in the metallic state. We report here the
experimental results of the RIXS measurements from the doped
La$_{1-x}$Sr$_x$MnO$_3$ which is compared with the previous result of
the undoped LaMnO$_3$ ($x=0$). The effect of the carrier doping on the
low energy excitations is focused by taking two representative
crystals ($x=0.2$ and $0.4$) and two different phases of either
insulator or metal as shown in Fig.\ \ref{fig:phase} of a simplified
phase diagram of La$_{1-x}$Sr$_x$MnO$_3$. In particular, the
temperature dependent features were elucidated in $x=0.2$ crystal,
where the phase transition from metal to insulator occurs associated
with the ferromagnetic transition governed by the double exchange
mechanism. The recent studies also suggest melting of the orbital long
range order at this MIT. Therefore it is worth to search the effect of
the orbital to the electronic excitations across the MIT boundary.

\section{Experimental}
We carried out the RIXS experiments at the beam line 11XU at
SPring-8. A spectrometer for inelastic x-ray scattering was installed
in this beam line\cite{Inami1}. Incident x-rays from a SPring-8
standard undulator were monochromatized by a diamond (111) double
crystal monochromator, and were focused on a sample by a horizontal
mirror. A Si (333) channel cut secondary monochromator was inserted
before the horizontal mirror when high energy resolution is
required. Horizontally scattered x-rays are analyzed in energy by a
diced and spherically bent Ge (531) crystal. The polarization vector
of incident x-rays and the scattering vector are in the horizontal
plane. The energy resolution in the experiments was 230 meV estimated
from the full width half maximum (FWHM) of quasielastic scattering,
when the Si (333) monochromator was used. The energy resolution
without the Si(333) monochromator is about 500 meV.  Single crystals
of La$_{0.6}$Sr$_{0.4}$MnO$_3$ and La$_{0.8}$Sr$_{0.2}$MnO$_3$ were
used.

\begin{figure}
\includegraphics[scale=0.45]{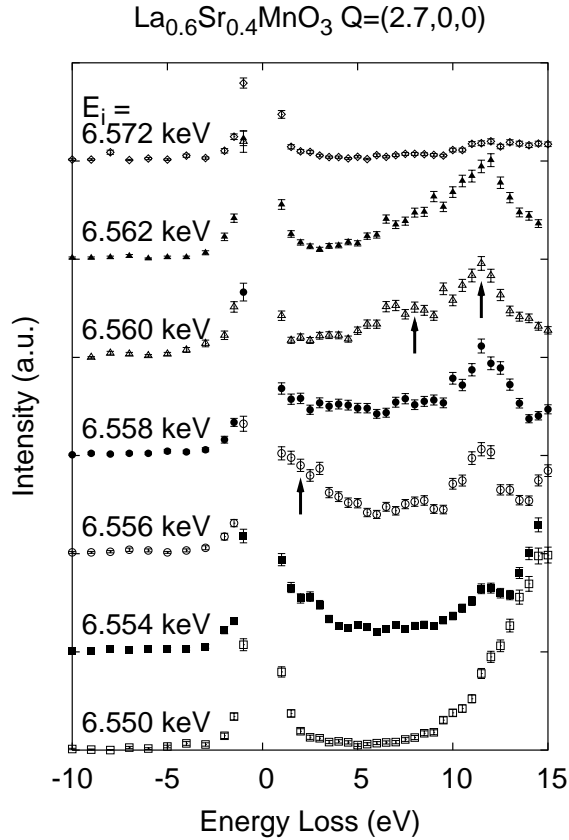}
\caption{Resonant inelastic x-ray scattering spectra of
La$_{0.6}$Sr$_{0.4}$MnO$_3$ as a function energy loss at some
representative incident x-ray energies ($E_i$). The energy resolution
is about 500 meV, and the scattering vector is fixed at
$\bm{Q}=(2.7,0,0)$. Three resonantly enhanced excitations are
indicated by the arrows. The strong intensity above 10 eV in the
spectrum of $E_i$ = 6.550 keV comes from the Mn $K\beta_5$ emission
line.}
\label{fig:x04eidep}
\end{figure}

The improvement of the energy resolution using the Si(333)
monochromator is crucially important in this study. We could observe
the peak feature of the excitation from the LHB to the UHB in high
resolution experiment (Fig.\ \ref{fig:xdep}(a)), while we could not
see it in the low energy resolution (Fig.\ \ref{fig:x04eidep}).

First we measured the spectra varying the energy of incident x-ray
($E_i$) at the fixed scattering vector $\bm{Q}=(2.7,0,0)$ to determine
a resonant energy. Though the crystal structure of
La$_{0.6}$Sr$_{0.4}$MnO$_3$ is rhombohedral\cite{Urushibara1}, we use
the index of $Pbnm$ orthorhombic notation to compare easily with our
previous paper\cite{Inami2}. All the data of this compound were taken
at room temperature. The results are shown in Fig.\
\ref{fig:x04eidep}. Three resonantly enhanced features can be seen at
the shoulder of the elastic peak, 8 eV, and 12 eV near the $K$
absorption edge of Mn. The resonant energy is slightly different in
each excitation. The scattering intensity of the excitation at the
shoulder of the elastic peak becomes strong at $E_i=6.556$ keV, while
those of 8 eV and 12 eV reach their maxima at higher $E_i$. The
difference of resonant energy between the excitations was also
observed in LaMnO$_3$\cite{Inami2} and La$_2$CuO$_4$ \cite{Kim1},
which indicates that the intermediate state involved in the excitation
is different. Hereafter the energy of incident x-ray was selected at
6.556 keV to focus mainly on the excitation at low energy.

\section{Results and discussion}
\subsection{Doping dependence of RIXS spectra}

\begin{figure}
\includegraphics[scale=0.45]{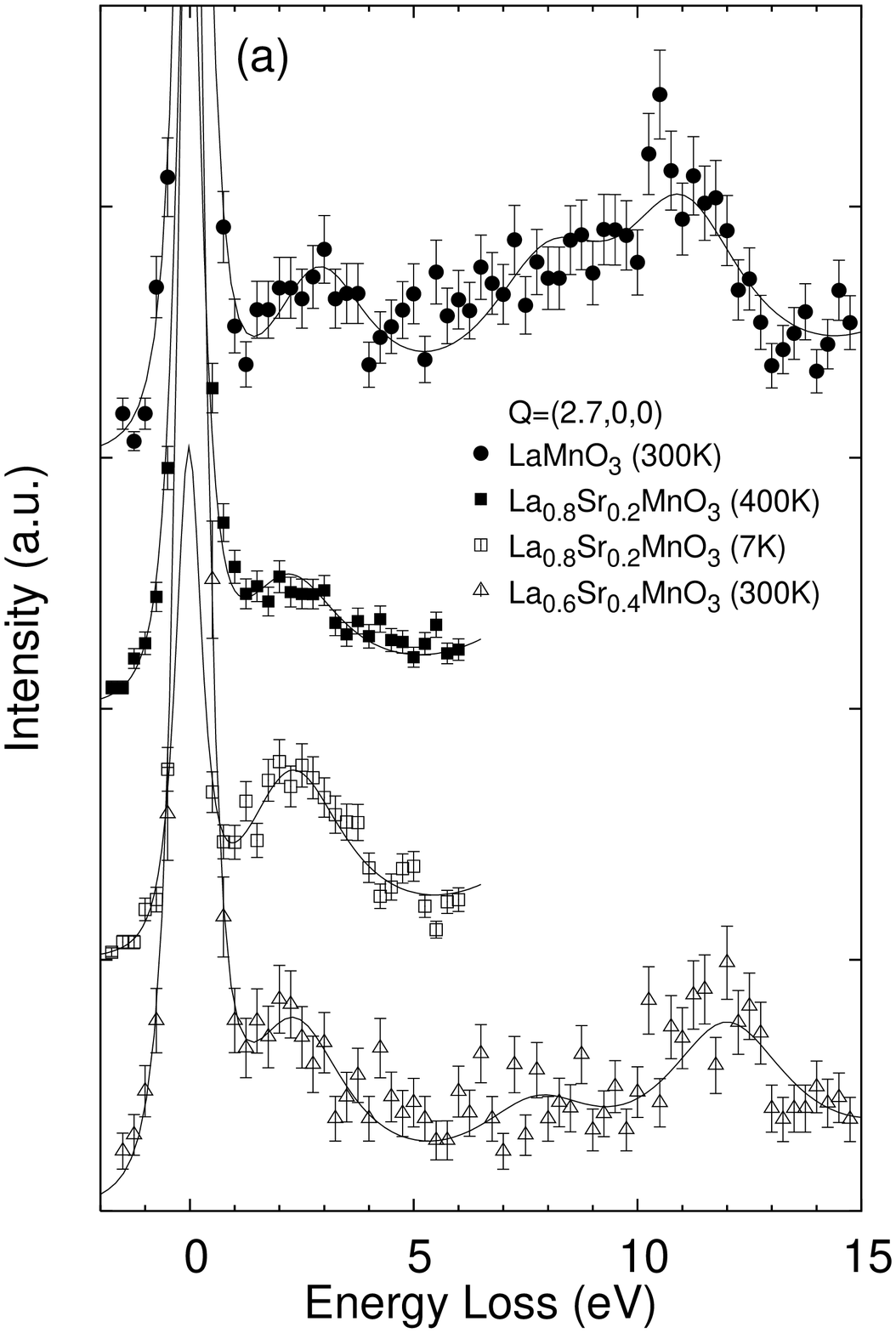}
\includegraphics[scale=0.45]{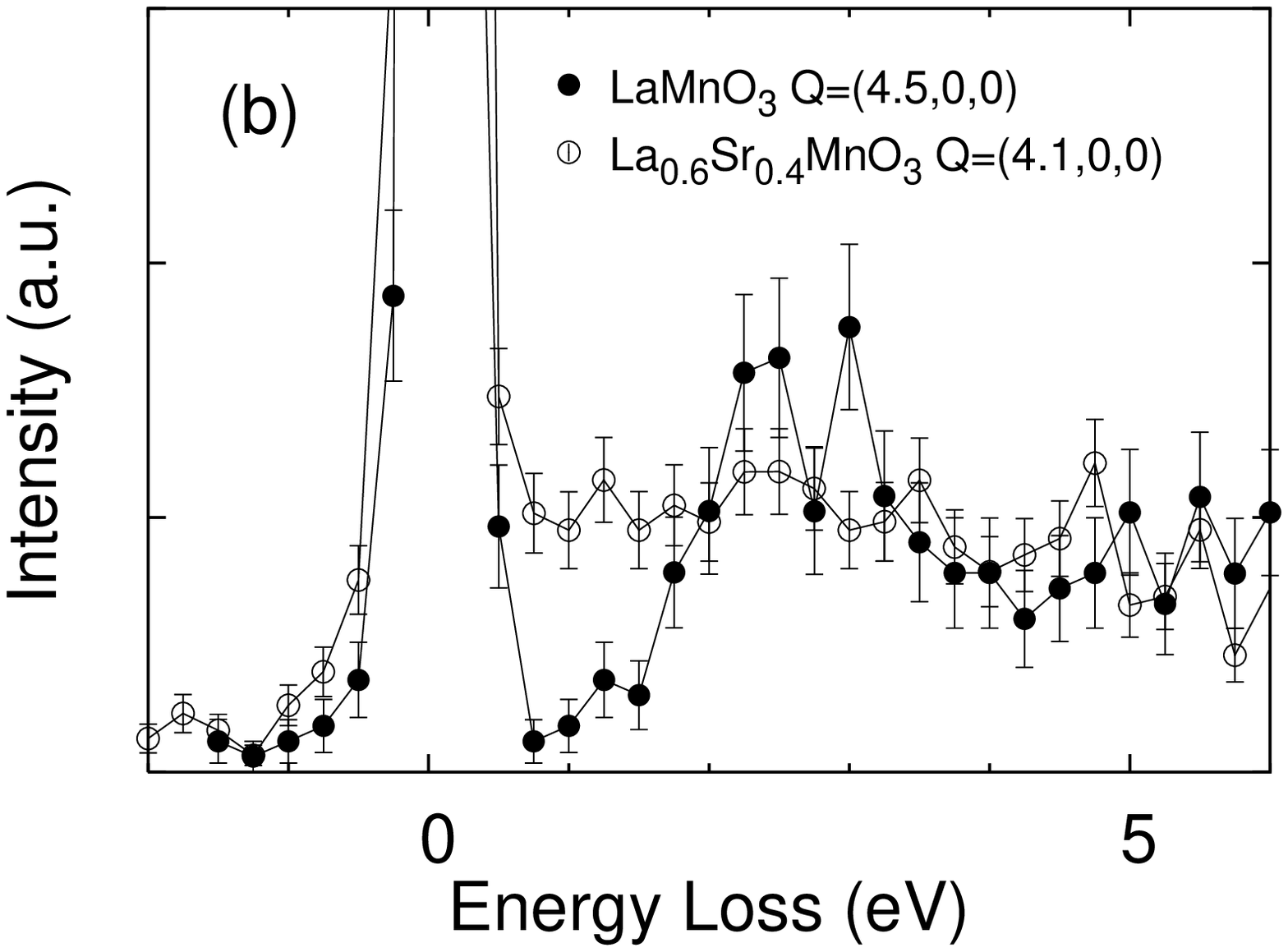}
\caption{Resonant inelastic x-ray scattering spectra of
La$_{1-x}$Sr$_x$MnO$_3$ ($x=0$, $0.2$, $0.4$). The energy resolution
is about 230 meV. (a) RIXS spectra at $\bm{Q}=(2.7,0,0)$. The solid
lines are fitting results assuming the elastic peak, three Lorentz
functions for excitations, and fluorescence (Mn-$K\beta_5$ line). (b)
The excitation in the low energy region measured at high scattering
angle. }
\label{fig:xdep}
\end{figure}

RIXS spectra of La$_{1-x}$Sr$_x$MnO$_3$ are shown in Fig.\
\ref{fig:xdep}(a), where each $x$ and $T$ corresponds to the filled
circle in Fig.\ \ref{fig:phase}. Upper two curves in the figure were
taken in the insulating phase, and lowers are those in the metallic
phase. It should be emphasized first that a salient peak commonly
appears at around 2 eV besides peaks at higher energies of 8 and 12
eV. The peak in the lowest energy can be assigned to be the electron
excitations from the LHB to the UHB as mentioned in the
introduction. Looking closely at the spectra, the spectral weight of
these peaks tends to shift toward lower energy by the hole-doping.
The experimental fact of the decrease of the energy gap between two
separated $e_g$ bands suggests the decrease of electron correlations
by the hole doping, which overcomes lower shift of Fermi energy in the
LHB.

Another important difference between two excitation spectra of
LaMnO$_3$ and La$_{0.6}$Sr$_{0.4}$MnO$_3$ is more clear in the data
taken at higher scattering angle ($2\theta$) of almost 90 degree. A
set of the high resolution spectra of LaMnO$_3$ and
La$_{0.6}$Sr$_{0.4}$MnO$_3$ are depicted in Fig.\ \ref{fig:xdep}(b).
The intensity of incoherent elastic scattering is nearly proportional
to $\cos^22\theta$ in our experimental conditions, namely, the
$\pi$-polarization of incident x-ray is used, and the elastic
scattering is weak at $2\theta\sim90$ degree. Since the momentum
dependence of the energy dispersion for the lowest excitations is
quite small, the difference of momentum between two spectra in Fig.\
\ref{fig:xdep}(b) is not important. A gap feature was clearly observed
in the spectrum of LaMnO$_3$, while the gap is partially filled in
La$_{0.6}$Sr$_{0.4}$MnO$_3$.

Even though the Hubbard gap is filled in La$_{0.6}$Sr$_{0.4}$MnO$_3$,
fairly large spectral weight remains at the excitation from the LHB to
the UHB, and forms the peak feature, as seen in Fig.\
\ref{fig:xdep}(a). It gives the direct evidence of the strongly
correlated electron nature.

It is natural that the origin of the excitations at 8 and 12 eV in
La$_{0.6}$Sr$_{0.4}$MnO$_3$ is similar to that in LaMnO$_3$. The
excitations at 8 and 11 eV in LaMnO$_3$ are the charge transfer (CT)
excitations from the O $2p$ orbitals to the Mn $3d$ and Mn $4s/4p$
orbitals, respectively.  In La$_{0.6}$Sr$_{0.4}$MnO$_3$ the former
corresponds to the excitation at 8 eV, and the latter is that at 12
eV. The peak position of the excitation from O $2p$ to the Mn $4s/4p$
is different between LaMnO$_3$ (11 eV) and La$_{0.6}$Sr$_{0.4}$MnO$_3$
(12 eV), as seen in Fig.\ \ref{fig:xdep}(a). Probably it comes from
different valence state of the manganese atom.  LaMnO$_3$ has only one
valence state of manganese atom (Mn$^{3+}$), while
La$_{0.6}$Sr$_{0.4}$MnO$_3$ has two (Mn$^{3+}$ and
Mn$^{4+}$). Interatomic distance between Mn$^{4+}$ and surrounding
oxygen is shorter than that between Mn$^{3+}$-O. The shrinkage of the
bond length makes the hybridization between O $2p$ and Mn $4p$
orbitals strong. The energy level of the Mn $4p$ orbital becomes
higher because it forms mainly an anti-bonding orbital, while the
energy level of O $2p$ bonding orbital becomes lower. As a result, the
excitation energy increases in the highly oxidized manganese.

\subsection{Momentum and azimuthal angle dependence of
La$_{0.6}$Sr$_{0.4}$MnO$_3$}

\begin{figure}
\includegraphics[scale=0.45]{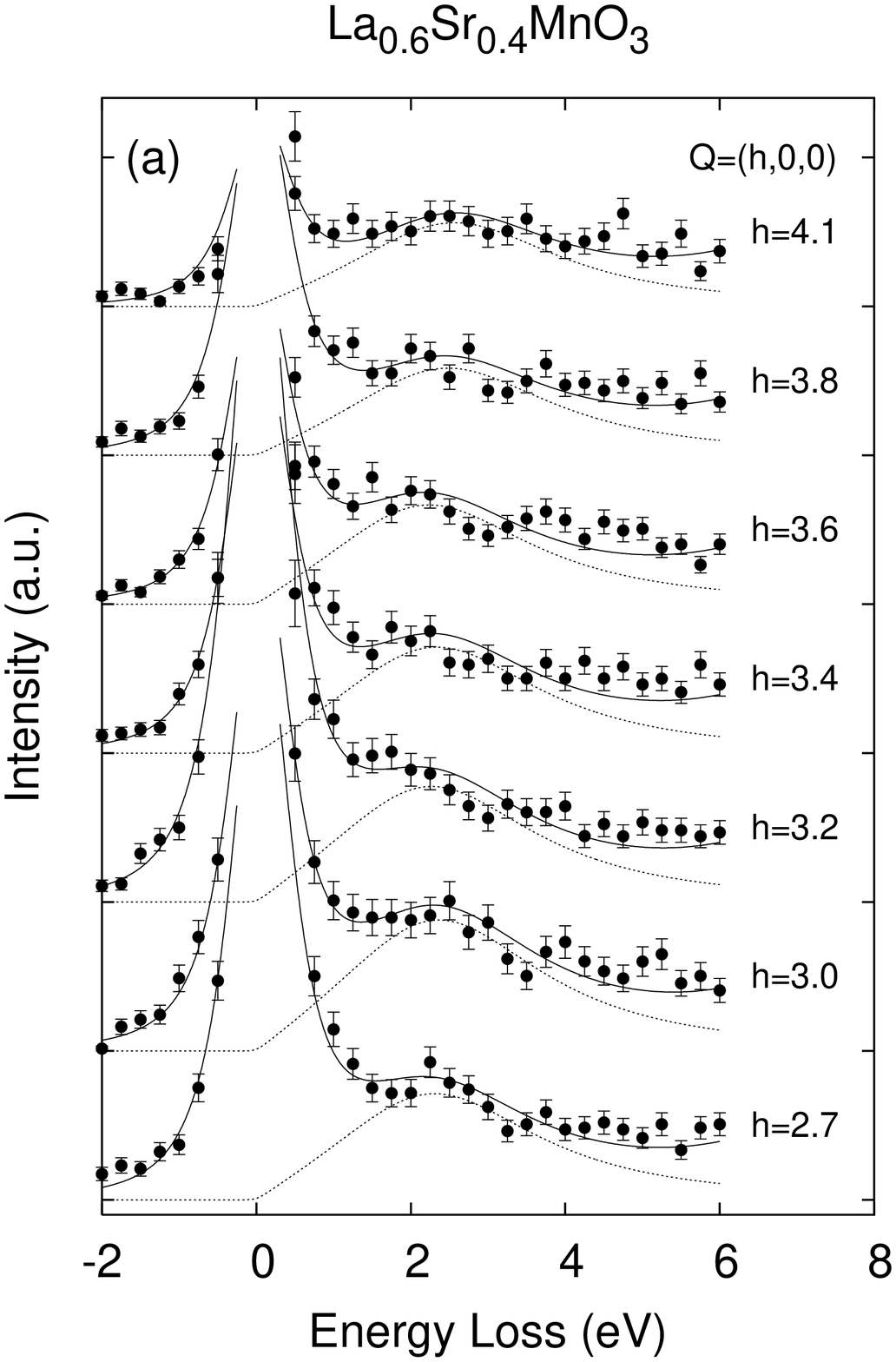}
\includegraphics[scale=0.45]{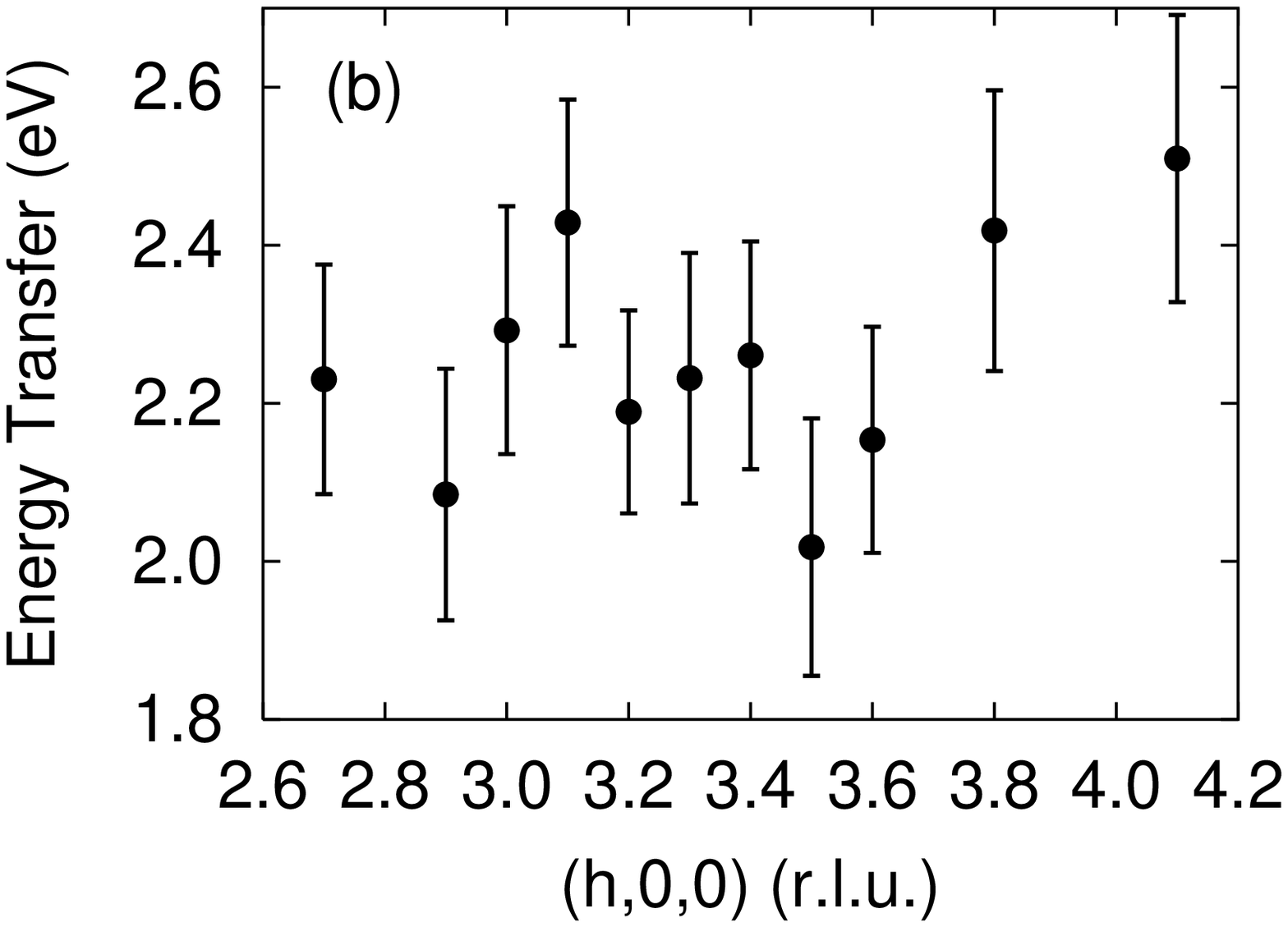}
\caption{(a) : The RIXS spectra of La$_{0.6}$Sr$_{0.4}$MnO$_3$ at
various scattering vectors. The energy resolution is about 230
meV. Solid circles are observed points. Solid and dashed lines are the
results of fitting of overall spectra and excitations from LHB to UHB,
respectively. (b) : Dispersion relation of the excitation from the LHB to
the UHB.}
\label{fig:x04qdep}
\end{figure}

The RIXS spectra of La$_{0.6}$Sr$_{0.4}$MnO$_3$ at various scattering
vectors are presented in Fig. \ref{fig:x04qdep}(a) where solid lines
show the result of the fitted curves. In orthorhombic notation of
La$_{0.6}$Sr$_{0.4}$MnO$_3$, $h=4$ and $h=3$ correspond to the
Brillouin zone center and the zone boundary, respectively.  The
momentum transfer dependence in the spectral shape was found to be
small. In order to elucidate the dispersion relation quantitatively,
we analyzed the observed data by fitting to the Lorentzian with the
fixed energy width of the excitations at 3 eV. The tail of the elastic
scattering or quasi-elastic component in the energy loss side (Stokes)
was evaluated from the energy gain side (anti-Stokes). The observed
intensities near 6 eV were considered to contain a tail from the peak
at 8 eV in part, and therefore the contribution from the peak at 8 eV
was taken into. The calculated spectral shape was fitted well as seen
in the figure. The obtained peak positions corresponding to the gap
energy between the LHB and the UHB are plotted in Fig.\
\ref{fig:x04qdep}(b). The magnitude of the dispersion is small and at
most 0.4-0.5 eV in total, which is comparable to that of LaMnO$_3$.

In general, the RIXS spectra are approximated as the convolution of
the occupied band and unoccupied band. In LaMnO$_3$ where the orbital
long-range order is realized, the occupied LHB band consists of the
orbital state represented by either $3d_{3x^2-r^2}$ or $3d_{3y^2-r^2}$
wave function which is alternately aligned. Thence the unoccupied
state should be $3d_{y^2-z^2}$/$3d_{z^2-x^2}$. As a result, hopping to
the nearest neighbor site in the UHB is forbidden in the $ab$ plane
because of the orthogonality of the wave functions. This is the reason
why the UHB of LaMnO$_3$ is narrow and the band dispersion is
flat. This fact also gives a weak dispersion relation in RIXS in
LaMnO$_3$. On the other hand, due to the fact of no orbital long range
order in La$_{0.6}$Sr$_{0.4}$MnO$_3$, hopping to the nearest neighbor
can be allowed in the UHB which gives rise to the larger dispersion
relation. The observed dispersion of the excitation spectrum of
La$_{0.6}$Sr$_{0.4}$MnO$_3$, however, is comparable to that of
LaMnO$_3$. This result can be possibly understood when the
ferromagnetic metal of La$_{1-x}$Sr$_x$MnO$_3$ has a short range
correlation of orbitals where the fluctuation of the orbital is enough
slow compared to the transition time of the x-rays. Slow fluctuation
of the orbital is regarded as a static disorder for the x-rays, and
also the excited electrons seem to be far from the band
electrons. Incoherent carrier motion in the crystal can be considered
as a kind of localization, and the weak dispersion observed in
La$_{0.6}$Sr$_{0.4}$MnO$_3$ is naturally comprehensible.

\begin{figure}
\includegraphics[scale=0.45]{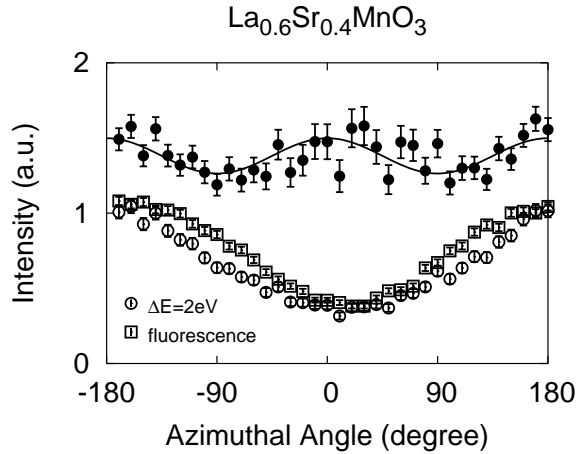}
\caption{Azimuthal angle ($\psi$) dependence of
La$_{0.6}$Sr$_{0.4}$MnO$_{3}$.  The open circles and squares are the
scattering intensity of the energy transfer 2 eV at $\bm{Q}=(2.7,0,0)$
and the fluorescence yield, respectively. The filled circles represent
the scattering intensity divided by the fluorescence. The solid lines
are the fitting result using a function of $A(1+B\cos^2\psi)$.}
\label{fig:azdep}
\end{figure}

This scenario of the short range orbital correlation was also
supported by another experiment of the polarization dependence of the
scattering intensity. We presented the azimuthal angle dependence of
the scattering intensity of 2 eV and the fluorescence as the reference
in Fig.\ \ref{fig:azdep}. Both show the oscillation of $2\pi$ period,
which probably comes from the difference between the angles of
incidence and reflection. When we divide the scattering intensity of 2
eV by the fluorescence yield, we can clearly see the two-fold
symmetry, which is reminiscent of the case of LaMnO$_3$. In the
LaMnO$_3$, the two-fold symmetry in the azimuthal angle dependence is
stronger and the characteristics of the orbital excitation from the
orbital ordered phase. We evaluated the ratio of the oscillating part
to the constant part using a function of $A(1+B\cos^2\psi)$, where
$\psi$ is the azimuthal angle. The origin of the $\psi$ is defined
when $c^{\ast}$ lies in the scattering plane. We obtained $B=0.38$ and
$0.19$ for LaMnO$_3$ and La$_{0.6}$Sr$_{0.4}$MnO$_{3}$, respectively.
The smaller value of $B$ for La$_{0.6}$Sr$_{0.4}$MnO$_{3}$ may be
attributed to the shorter correlation length.

\subsection{Temperature dependence of La$_{0.8}$Sr$_{0.2}$MnO$_3$}

\begin{figure*}
\includegraphics[scale=0.7,angle=-90]{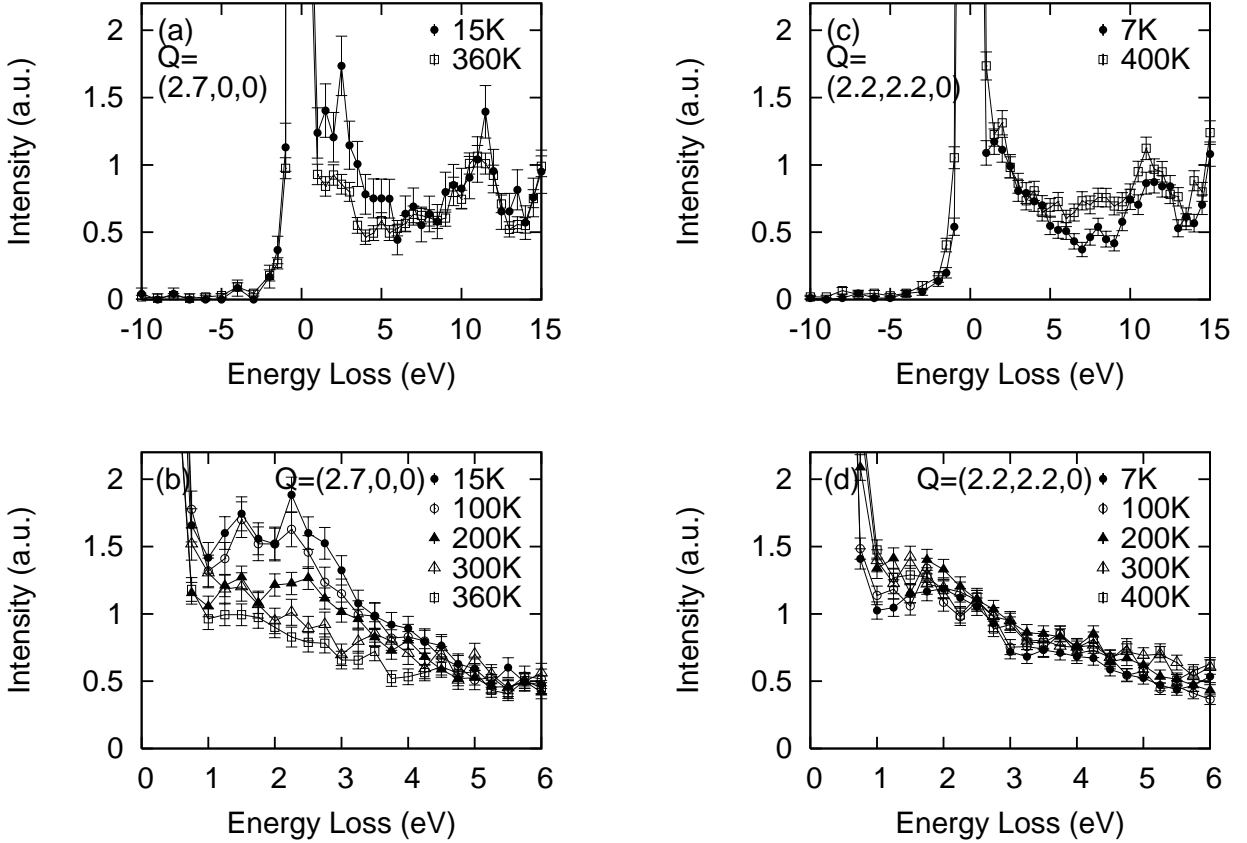}
\caption{RIXS spectra of La$_{0.8}$Sr$_{0.2}$MnO$_3$ at some
  temperatures. The energy resolution is about 500 meV. The scattering
vectors are $\bm{Q}=(2.7,0,0)$ for (a) and (b), and
$\bm{Q}=(2.2,2.2,0)$ for (c) and (d).}
\label{fig:x02tdep}
\end{figure*}

In this section we focus on the result of La$_{0.8}$Sr$_{0.2}$MnO$_3$
which shows the MIT at 309 K, associated with the ferromagnetic phase
transition \cite{Urushibara1}. It also shows a structural phase
transition at 100 K from rhombohedral to
orthorhombic\cite{Kawano1}. We measured the RIXS spectra across two
transition temperatures, as indicated by the arrow in Fig.\
\ref{fig:phase}.

The RIXS spectra of La$_{0.8}$Sr$_{0.2}$MnO$_3$ measured at several
representative temperatures are shown in Figs.\ \ref{fig:x02tdep}. The
spectra are similar to those of La$_{0.6}$Sr$_{0.4}$MnO$_3$ in Fig.\
\ref{fig:x04eidep}; there are three excitations at 2 eV, 8 eV, and
11.5 eV. The scattering intensity at 2-4 eV increases with decreasing
temperature at $\bm{Q}=(2.7,0,0)$ (Figs.\ \ref{fig:x02tdep}(a) and
(b)). In contrast, the intensity of $\bm{Q}=(2.2,2.2,0)$ is nearly
independent of temperature (Figs.\ \ref{fig:x02tdep}(c) and (d)). It
should be noted that the excitations at 11.5 eV is independent of
temperature in both scattering vectors. The spectra of $\bm{Q}=(3.3,0,0)$ were
confirmed to show similar temperature dependence to those of
$\bm{Q}=(2.7,0,0)$, and the spectra of $\bm{Q}=(2.6,2.6,0)$ is independent 
of temperature. We emphasize here the experimental fact that 
the temperature dependence in intensity depends on the direction of
scattering vector: in other words, the temperature dependence of the
inter-band excitations from the LHB to the UHB is quite anisotropic.

The temperature dependence of RIXS intensity can be compared with that
of the optical conductivity. The spectral weight of the optical
conductivity in La$_{0.8}$Sr$_{0.2}$MnO$_3$ shifts gradually from 1-2
eV to lower energy ($<1$ eV) with decreasing
temperature\cite{Saitoh1}. The strength of the optical conductivity at
around 2 eV decreases with decreasing temperature. Surprisingly, the
temperature dependence of the RIXS intensities of $\bm{Q}=(2.7,0,0)$
is opposite to that of the optical conductivity ($\bm{Q}=\bm{0}$),
namely, the intensities along $\langle h00 \rangle$ direction increase
with decreasing temperature.  Furthermore, the RIXS intensities show no
obvious temperature dependence along $\langle hh0 \rangle$. Since the
RIXS can measure the excitation at the finite momentum transfer, the
electronic excitations of La$_{0.8}$Sr$_{0.2}$MnO$_3$ should be
correctly understood by taking account of momentum dependence. The
qualitatively different temperature dependence just suggests the
influence of the orbital to the electronic conductivity, which has
often been discussed.

\begin{figure}
\includegraphics[scale=0.45]{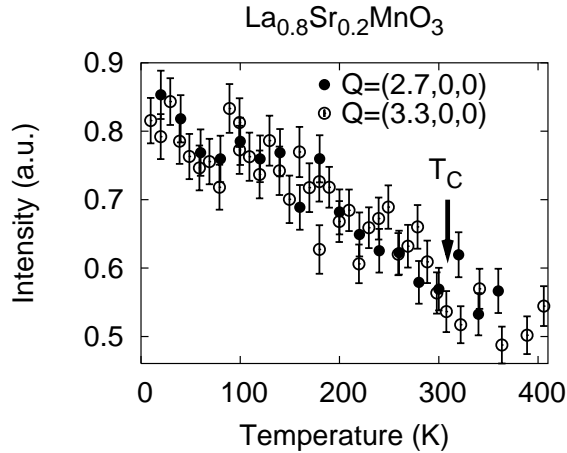}
\caption{Temperature dependence of the RIXS intensity at fixed energy
transfer of 2.25 eV in La$_{0.8}$Sr$_{0.2}$MnO$_3$.  The arrow indicates
the ferromagnetic transition temperature ($T_C$).}
\label{fig:tscan}
\end{figure}

A key to understand the temperature dependence as well as the
anisotropy of RIXS intensity is the transfer of electrons near the
Fermi energy. The double exchange interaction plays an essential role
for the stabilization of the ferromagnetic long range order in this
hole doped La$_{1-x}$Sr$_x$MnO$_3$ compounds. Because the probability
of the spin-flip excitations in the RIXS process is much smaller than
that of the spin-non-flip excitations, the ferromagnetic spin
order of $e_g$ electrons increases the transition probability from
the LHB to the UHB. In this sense, the change in RIXS intensity along
$\langle h00 \rangle$ may be attributed to the evolution of double
exchange ferromagnetic interaction. The temperature dependence of
scattering intensity of 2.25 eV at $\bm{Q}=(2.7,0,0)$ and $(3.3,0,0)$
in Fig.\ \ref{fig:tscan}, which qualitatively accords to that of the
bulk magnetization\cite{Urushibara1}: the intensity begins to increase
at $T_C$, and saturates around 150 K. On the other hand, the
scattering intensity is independent of temperature along $\langle hh0
\rangle$ (Mn-O-Mn direction), which might be reflected from the
ferromagnetic super-exchange interaction which is active in both
metallic and insulator phases.

The present results demonstrate that the temperature dependence of
the electronic excitation is quite anisotropic in
La$_{0.8}$Sr$_{0.2}$MnO$_3$, and the RIXS is a powerful technique to
study electronic excitation, especially in an anisotropic system,
because it can provide a plenty information of the electronic
excitation with {\it finite} momentum transfer.

\section{Conclusion}
In conclusion, we have measured resonant inelastic x-ray scattering
spectra of hole-doped manganese oxides, La$_{0.8}$Sr$_{0.2}$MnO$_3$
and La$_{0.6}$Sr$_{0.4}$MnO$_3$, and studied electronic excitations
across the Mott-Hubbard gap near the Fermi energy.

The salient peak structure which corresponds to the excitation from
the LHB to the UHB was observed in both La$_{0.8}$Sr$_{0.2}$MnO$_3$
and La$_{0.6}$Sr$_{0.4}$MnO$_3$, and the peak position shifts to lower
energy than that of LaMnO$_3$. Furthermore, the spectral weight
extended to lower energies than the Mott-Hubbard gap in metallic
La$_{0.6}$Sr$_{0.4}$MnO$_3$, though the peak of inter-band excitation
like LaMnO$_3$ still maintains. This is the first RIXS experiments in
which some characteristics of the strong electron correlation are
elucidated in the metallic state.  The excitations from the LHB to the
UHB in La$_{0.6}$Sr$_{0.4}$MnO$_3$ shows weak dependence on the
momentum transfer, and the magnitude of the dispersion is 0.4-0.5 eV.
The scattering intensity contains a component of the two-fold symmetry
in the azimuthal angle dependence, though no static orbital order
appears. All these characteristics of the inter-band excitations from
the LHB to the UHB indicate that the local correlation effect is
strong even in La$_{0.6}$Sr$_{0.4}$MnO$_3$.

In La$_{0.8}$Sr$_{0.2}$MnO$_3$, we found a clear temperature
dependence in intensity around 2-4 eV. The RIXS intensity of
$\bm{Q}=(2.7,0,0)$ increases with decreasing temperature, while the
scattering intensity of $\bm{Q}=(2.2,2.2,0)$ is independent on
temperature. It is a remarkable contrast with the optical
conductivity, whose spectral weight at 2 eV decreases with decreasing
temperature. The anisotropic temperature dependence might be
attributed to an anisotropy of the ferromagnetic exchange interaction.
Even though the energy dispersion is small in La$_{1-x}$Sr$_x$MnO$_3$,
we could demonstrate a momentum dependent feature of the electronic
excitations by the RIXS technique for the first time in the sense that
the temperature dependence of the scattering intensity is
anisotropic. Furthermore, the importance and usefulness of the RIXS
measurements are recognized throughout the present studies because the
momentum dependent electronic excitations in the manganite systems
reveal the mutual interactions between charge, spin and orbital.

\begin{acknowledgments}
We acknowledge to Dr. P. Abbamonte for making the Ge analyzer.  This
work was supported financially in part by Core Research for
Evolutional Science and Technology (CREST), sponsored by the Agency of
the Japan Science and Technology.
\end{acknowledgments}

\bibliography{ixs,mangane}

\end{document}